\documentclass{aa}  

\usepackage{graphicx}
\usepackage{amsmath}
\usepackage{amssymb}
\usepackage{natbib} 
\usepackage{txfonts}
\usepackage{comment}
\usepackage{xcolor}

\usepackage{subfigure}

\usepackage{hyperref}
\hypersetup{
    colorlinks=true,
    linkcolor=blue,    
    urlcolor=blue,
    citecolor= blue
    }
    
\begin{document} 
   \title{\texttt{SAGE:} A tool to constrain impacts of stellar activity on transmission spectroscopy}

   \author{H. Chakraborty
          \and
          M. Lendl
          \and
          B. Akinsanmi
          \and
          D.J.M. Petit dit de la Roche
          \and
          A. Deline
          }

  \institute{ Geneva Observatory, University of Geneva, Chemin Pegasi 51, 1290 Versoix, Switzerland\\
              \email{Hritam.Chakraborty@unige.ch}
             \thanks{GitHub: \url{https://github.com/chakrah/sage}}
            }


  \abstract{

Transmission spectroscopy is a proven technique to study a transiting exoplanet's atmosphere. However, stellar surface inhomogeneities, spots and faculae, alter the observed transmission spectra: the stellar contamination effect. The variable nature of the stellar activity also makes it difficult to stitch together multi-epoch observations and evaluate any potential variability in the exoplanet’s atmosphere. This paper introduces \texttt{SAGE}, a tool to correct for the time-dependent impact of stellar activity on transmission spectra. It uses a pixelation approach to model the stellar surface with spots and faculae, while fully accounting for limb-darkening and rotational line-broadening. The current version is designed for low to medium-resolution spectra. We used \texttt{SAGE} to evaluate stellar contamination for F to M-type hosts, testing various spot sizes and locations, and quantify the impact of limb-darkening. We find that limb-darkening enhances the importance of the spot location on the stellar disk, with spots close to the disk center impacting the transmission spectra more strongly than spots near the limb.  Moreover, due to the chromaticity of limb darkening, the shape of the contamination spectrum is also altered. Additionally, \texttt{SAGE} can be used to retrieve the properties and distribution of active regions on the stellar surface from photometric monitoring. We demonstrate this for WASP-69 using TESS data, finding that two spots at mid-latitudes and a combined coverage fraction of $\sim$1\% are favoured. \texttt{SAGE} allows us to connect the photometric variability to the stellar contamination of transmission spectra, enhancing our ability to jointly interpret transmission spectra obtained at different epochs.

} 
  

   \maketitle
%

\section{Introduction}

Transmission spectroscopy is a powerful tool to study the atmospheres of transiting exoplanets. The method is based on the property that the periodic dimming of a star during the planetary transit has a colour dependence \citep{2000ApJ...537..916S}. If the atmosphere of a planet is more opaque at certain wavelengths due to absorption by atomic or molecular species, then it will result in a larger flux drop, i.e., a deeper transit. By studying these variations in the transit depth at different wavelengths, we can derive the atmospheric properties of the planet. The method has been used to detect atmospheric features such as Rayleigh scattering (\citealp{2008A&A...485..865L}, \citealp{2017MNRAS.467.4591G}), clouds (\citealp{2016A&A...587A..67L}, \citealp{2020AJ....160...51A}), hazes (\citealp{2008MNRAS.385..109P}, \citealp{2015MNRAS.447..463N}, \citealp{2016Natur.529...59S}) and also different atomic and molecular species such as Na \citep{2008A&A...487..357S, 2015A&A...577A..62W}, K \citep{2021A&A...645A..24B}, Fe \citep{2019A&A...627A.165H, 2020Natur.580..597E}, He \citep{2018Sci...362.1384A}, H$_{2}$O, CO$_{2}$ (\citealp{2023Natur.614..649J}) and TiO/ VO (\citealp{2016ApJ...822L...4E})  in the atmospheres of exoplanets. 
Transmission spectroscopy involves measuring small variations in the transit depth. Thus, it is often desirable to study planets with deeper transits as they produce larger amplitude signals. In the case of small Earth-like planets, we are limited to planets revolving around small stars as the transit depth scales with the square of the ratio between the planetary and stellar radii; but these small stars are often magnetically active \citep{2012LRSP....9....1R, 2016PhR...663....1S}. This activity manifests itself as surface inhomogeneities in the form of cooler (spots) and hotter (faculae) regions on the photosphere, and these impact the measurement of accurate transit depths and thus the transmission spectra. 

The impact of activity features such as spots and faculae has been studied by multiple authors on different planetary systems. Their effect on the transit light curves can be broadly classified into two types; direct and indirect. The direct effect corresponds to the event in which an active region is located on the transit chord during the planetary transit. When the planet occults the active region, it results in anomalies in the transit light curves (\citealp{2003ApJ...585L.147S}, \citealp{2011MNRAS.416.1443S}). In the case of spots, it leads to a brief rise in flux that can lead to an underestimation of the transit depth, which can be as high as 10\% at wavelengths of $\sim$ 500 nm (\citealp{2014A&A...568A..99O}). In the case of faculae, the transit depth is overestimated. This effect can be mitigated by using routines that model the occulted active regions along with the transit light curve such as \texttt{PyTranSpot} (\citealp{2018A&A...610A..15J}), \texttt{SOAP-T} (\citealp{2013A&A...549A..35O}),  or even Gaussian Process (\citealp{2018AJ....156..124B}). 
   
The indirect effect of active regions on transmission spectra arises from the fact that the measurement of transit depth is differential. Transit depth is measured as the difference between the in-transit and out-of-transit flux of the star. As a transiting planet blocks light from the narrow transit chord of the star, the inhomogeneities on the stellar surface cause the total disc-integrated flux (i.e. the in-transit or the out-of-transit flux) to differ from the flux blocked from below the transit chord. Thus, altering the transit depth measurements. In addition, due to spectral differences in the emerging flux, the effect is wavelength-dependent. This phenomenon is commonly referred to as stellar contamination of transmission spectra. In the case of cooler spots, possessing a redder spectrum than the disk-averaged stellar light, we observe a rise in transit depth at bluer wavelengths, mimicking the effect of Rayleigh scattering in the planetary atmospheres. For the highly active transiting system HD189733, the predicted difference in transit depth between ultra-violet and infrared wavelengths is $\sim$800 ppm \citep{2008MNRAS.385..109P, 2013MNRAS.432.2917P, 2014ApJ...791...55M}. 

To accurately determine the extent of stellar contamination on transmission spectra, accurate estimates of the physical properties of active regions such as their temperature, size and location on the stellar surface during the transit are needed. One approach to constraining these properties is to photometrically monitor the star throughout at least a full rotation period around the transit observation, and use the amplitude and shape of the variability to retrieve the properties of the active regions present on the stellar surface \citep{2007A&A...476.1347P, 2008MNRAS.385..109P}. \cite{2017ApJ...834..151R,2018ApJ...853..122R} presented a forward-modelling approach to link the spot/facula coverage fractions to the amplitude of the variability in the Kepler light curves, and use those to predict the stellar contamination spectra. While allowing an order-of-magnitude evaluation of the effect, this technique is limited in its application to specific systems and observations by the degeneracy in the active region's size and temperature, and its ignorance of the exact placement of the spots on the stellar disk. An alternative approach by \cite{2017ApJ...844...27Z} utilises the out-of-transit spectrum compared to the spectrum of the star during an inactive phase to calculate the stellar contamination without any prior measurements of the variability or the configuration of active regions. This technique hinges on the precision of the spectra and requires the availability of spectra of the \textit{quiet} star, which may often be complicated to secure.

The magnitude of stellar contamination on the transmission spectrum has an epoch dependence due to the evolution of active regions. This evolution, and the variations associated with stellar rotation, pose a significant problem when combining or comparing multi-epoch observations. As transmission spectroscopy becomes increasingly precise with the onset of the JWST era, there is a need to carefully disentangle the effects of stellar contamination from any effects originating in the atmospheres of exoplanets.  

In this work, we introduce the Stellar Activity Grid for Exoplanets (\texttt{SAGE}), a tool to constrain and correct for the epoch-dependent impact of active regions on transmission spectra. The code calculates the stellar contamination from a simulated disc-resolved stellar spectrum with known spot properties. This approach allows us to study additional effects such as limb-darkening, which critically shapes the stellar contamination. It can also simulate the flux variability associated with the stellar rotation for a given spot configuration and can be used in a Markov chain Monte Carlo (MCMC) framework to retrieve spot properties from rotational variability measurements. 

This paper is structured as follows: in Section 2, we present the astrophysical model that we used to model  spots on the stellar photosphere. We also present how the variability and contamination models are calculated in \texttt{SAGE}. In Section 3, we discuss the effects of the inclusion of limb-darkening and the projection effects of spots on the strength of stellar contamination. In Section 4, we use the variability model to draw inferences about the surface inhomogeneities of WASP-69. Sections 5 and 6 present the discussion, summary and conclusion of our study.

\section{Modelling inhomogeneities of stellar photospheres}

Multiple routines, notably \texttt{SOAP-T} \citep{2013A&A...549A..35O}, \texttt{Ksint} \citep{2014MNRAS.444.1721M}, \texttt{PRISM}  \citep{2013MNRAS.428.3671T, 2015MNRAS.450.1760T}, \texttt{StarSim} \citep{2016A&A...586A.131H}, \texttt{ellc} \citep{2016A&A...591A.111M} and \texttt{PyTranSpot} \citep{2018A&A...610A..15J}, are available to model and correct for the "direct" effects of stellar activity on the transit light curves (star spot crossings). With \texttt{SAGE}, we introduce a Python tool to correct for the "indirect" effects of stellar activity on transmission spectra (stellar contamination). To do so, the code is equipped with two capabilities: i) it can fit the stellar rotational variability, constraining the distribution and size of active regions on the star, and ii) based on the derived distribution, it can compute the stellar contamination at any given rotational phase of the star.



With \texttt{SAGE}, the user creates a model star projected on a 2-dimensional pixel grid. In the model, the active regions on the stellar surface are parameterised using their number, position, size and temperature. Each pixel in this grid emits a spectrum whose properties are dependent on the local properties of the stellar surface, e.g., the pixels associated with cooler regions emit a cooler spectrum compared to the pixels associated with hotter regions. The emitted spectra are parameterised using the effective temperature, surface gravity and metallicity of the stellar photosphere. Additionally, stellar effects such as limb-darkening, rotational broadening and inclination can be modelled. By rotating this model star and integrating the emitted flux over a desired wavelength band, we obtain the flux  variability  for a given configuration of active regions.



\subsection{Model setup} 
\label{Section 2.1}
\begin{figure*}
  \includegraphics[width=17 cm]{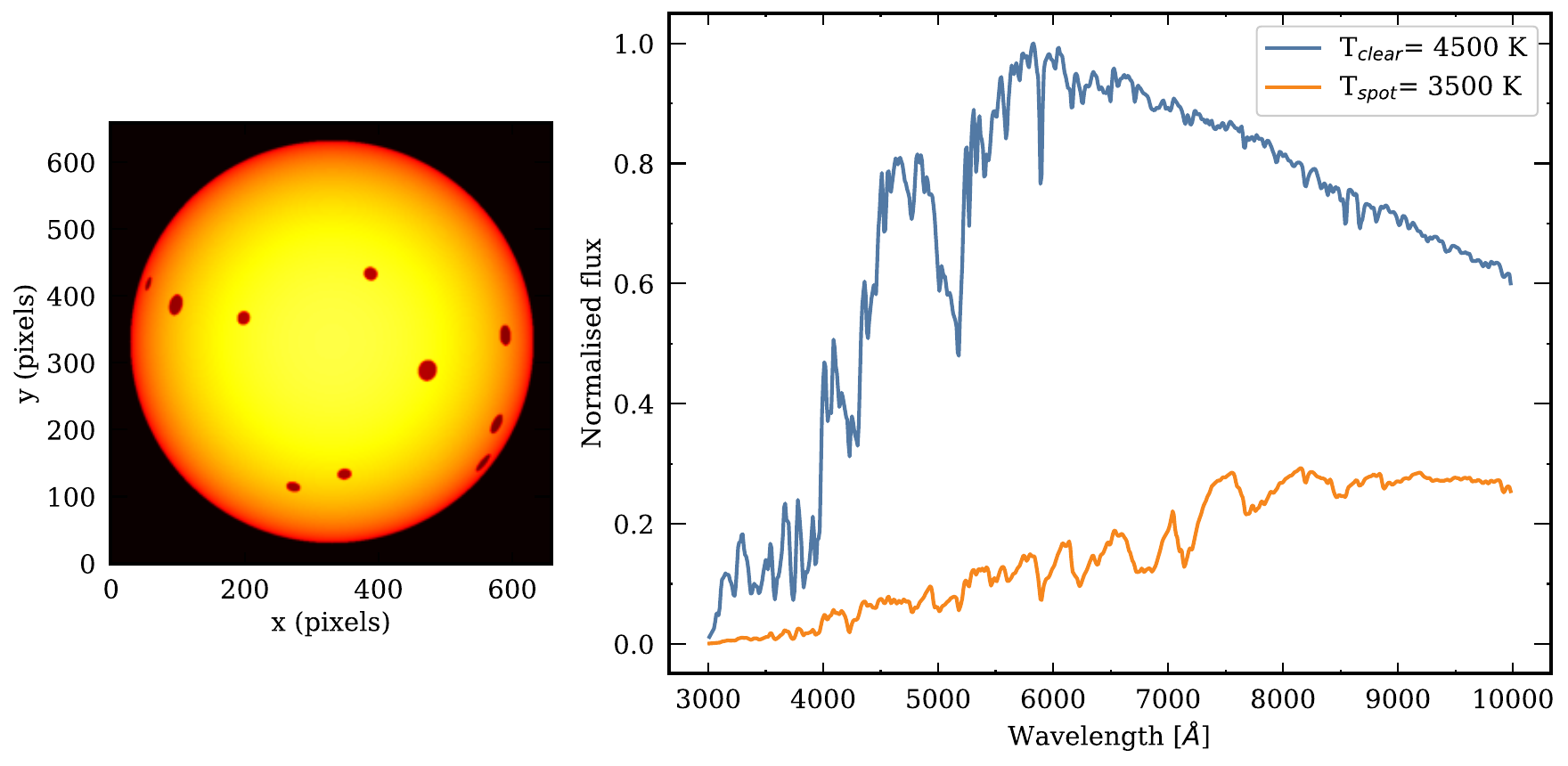}
  \caption{ \textit{Left:} A stellar sphere with LD and spots projected onto a 2-dimensional Cartesian pixel grid. \textit{Right:} The normalised model spectra used, obtained from the PHOENIX library. For the clear photosphere and for spots, we used temperatures of 4500 K and 3500 K, respectively. 
  }
  \label{Stellar spectrum}
\end{figure*}

\texttt{SAGE} uses a pixelation approach similar to that used in \texttt{PyTranSpot} \citep{2018A&A...610A..15J} and \texttt{PRISM} \citep{2013MNRAS.428.3671T, 2015MNRAS.450.1760T}. We project the 3-dimensional stellar sphere onto a 2-dimensional Cartesian grid of pixels each with the same projected area. This method allows to simultaneously model multiple components, such as spots, faculae, limb-darkening, etc. The size of the pixel grid ($n \times n$) is defined through the star-planet radius ratio together with a user-specified planet size in pixels. A higher planet pixel size provides a better spatial resolution of the inhomogeneities, but at the cost of computation time. A planet radius of 15 to 50 pixels is recommended when modelling occulted spots in transit light curves and we recommend using these limits for \texttt{SAGE} as well (\citealp{2018A&A...610A..15J}, \citealp{2015MNRAS.450.1760T}). The choice is also dependent on the resolution of the input spectra; a finer grid will provide better estimates of the effects of LD and rotational broadening relevant at high resolution. However, an increased resolution of spectra and grid comes with a higher required computation time.

We model surface inhomogeneities as homogeneous circular regions on the stellar disc. These circular regions deform to ellipses as regions move closer to the limb of the star. This deformation inherently corrects for projection effects that are especially relevant for active regions located near the limb of the star (for example, polar spots). From solar observations, we know that spots and faculae have complicated structures and often appear in groups \citep{2003A&ARv..11..153S, 2020A&A...644A..43P, 2021A&A...652A...9M}. However, we are currently limited in resolving such structures in other stars given the limited information available from spot occultations during transit.

We choose a spherical coordinate system with its origin located at the center of the stellar sphere. The inhomogeneities on the stellar photosphere are parameterised using their position, size and temperature/contrast with respect to the immaculate photosphere. These are modelled in SAGE using the following parameters:

\begin{itemize}
    \item \textit{Longitude} ($\theta$) varies between -180$^{\circ}$ and 180$^{\circ}$. The visible stellar disc ranges from -90$^{\circ}$ to 90$^{\circ}$ in the counter-clockwise direction. 
    
    \item \textit{Latitude} ($\phi$) varies between 90$^{\circ}$ and -90$^{\circ}$. The north pole of the star is at 90$^{\circ}$ and the south pole is at -90$^{\circ}$.
    
    \item \textit{Size} ($\alpha$) can be defined either as the angular size on the 3-dimensional stellar sphere or in terms of the absolute filling factor. The absolute filling factor is defined as the ratio between the surface area of the active region and the total stellar surface. 
    
    \item \textit{Contrast} ($\rho_{\rm{spot}}$) is the ratio between the intensities of an active region to the immaculate photosphere, disregarding limb darkening. It is a wavelength-dependent quantity that ranges from 0 to 1 for cooler regions and is $>$ 1 for hotter regions. 
\end{itemize}

\subsection{Stellar spectrum}

A model spectrum is allocated to each pixel on the stellar grid. Various spectral libraries may be chosen and we have tested both low/mid resolution such as ATLAS9 \citep{2003IAUS..210P.A20C} and PHOENIX \citep{2013A&A...553A...6H} and high-resolution spectral libraries such as Coelho \citep{2014MNRAS.440.1027C}. 
To account for the difference in the emergent spectrum between the active regions and clear photosphere, different model spectra are chosen for both clear ($F_{\rm{Clear}},_{\lambda}$) and active regions ($F_{\rm{Active}},_{\lambda}$). The model spectra are normalised to the peak of the clear spectrum before being embedded in the pixel grid (Fig. \ref{Stellar spectrum}). This doesn't alter the shape of the input spectrum but makes the computation more tractable. The calculation of the disc-integrated spectrum at high resolution requires the addition of essential stellar effects such as convective blueshift, which will be part of future work. 

The normalised disc-integrated spectrum of the star is calculated by summing the individual flux contributions from each pixel,

\begin{equation}
    \textrm{Flux}_{ \lambda} = \frac{\sum_{i}^{N_{\rm{active}}} F_{\rm{active}},_{ \lambda},_{ i} +  \sum_{j}^{N - N_{\rm{active}}} F_{\rm{clear}},_{\lambda},_{j}}{N} .
    \label{Equation: 1}
\end{equation}
    
\noindent Here, F$_{\rm{active, i}}$  is the flux emission from the $i^{\textrm{th}}$ pixel in the active regions and F$_{\rm{clear, j}}$ is the flux from the $j^{\rm{th}}$ pixel of the clear photospheres. $\rm{N}$ and $\rm{N}_{\rm{active}}$ are the total numbers of pixels  on the stellar disc and the  active regions, respectively. The wavelength coverage and resolution of the disc-integrated spectrum are the same as the input spectra. If the spectral resolution of the clear and active photosphere models do not match, then the spectrum with the higher resolution is binned down to the lower resolution spectrum.
  
\subsection{Rotational broadening and limb-darkening effects}

One of the main advantages of using a pixelation approach is the possibility to incorporate effects such as rotational broadening, LD and gravity-darkening in the stellar grid. The magnitude of these effects varies based on the properties of the star. The current version of \texttt{SAGE} is set up to include both, LD and rotational-broadening effects. 

In our model, we assume the star is a rigid rotator to calculate the effect of rotational broadening. The model spectrum for each pixel is shifted in wavelength according to its radial velocity, which depends on the pixel location and the rotational velocity of the star. The injected model spectra should be free from the effects of stellar rotation. The rotation of the star in our model is parameterised using the equatorial velocity of the star in km s$^{-1}$. The resolution of the input spectra puts a natural limit on the velocity resolution, and thus the effects of rotational broadening can only be studied with sufficiently high-resolution input spectra.



To model LD, we choose the quadratic LD law:

\begin{equation}
    \frac{I(\mu)}{I(1)} = 1 - u_{1} (1 - \mu) - u_{2} (1 - \mu)^{2} ,
\end{equation}

\noindent where $u_{1}$ and $u_{2}$ are the wavelength-dependent LD coefficients, I(1) is the intensity at the center of the stellar disc and $\mu$ = cos($\theta$), where $\theta$ is the angle between the line of sight and normal to the stellar surface.

In \texttt{SAGE}, the required LD coefficients, $u_{1}$ and $u_{2}$, are calculated using the Limb-Darkening Coefficients and Uncertainties (\texttt{LDCU}) package \footnote{\url{https://github.com/delinea/LDCU}}. \texttt{LDCU}  is a modified version of the Python routine by \cite{2015MNRAS.450.1879E} that computes the LD coefficients and their corresponding uncertainties using a set of stellar intensity profiles accounting for the uncertainties on the stellar parameters. The stellar intensity profiles are generated based on two libraries of synthetic stellar spectra: ATLAS \citep{1979ApJS...40....1K} and PHOENIX \citep{2013A&A...553A...6H}. The LD coefficients in \texttt{LDCU} are parameterised by the effective temperature ($T_{\rm{eff}}$), surface gravity ($log$ $g$), metallicity ($[M/H]$), and turbulent velocity ($v_{\rm{tur}}$) of the star. 

LD can be different for quiet and active regions. From Solar observations, \cite{2004A&A...419..735C} measured a much stronger LD for a spotted region compared to quiet regions at mid-ultraviolet wavelengths (175-210 nm). For faculae on the Sun, surface brightness increases towards the limb \citep{1999ApJ...518..480F, 1977ApJ...217..657C, 2010A&A...512A..39M}. Unfortunately, the LD behaviour of active regions has not been studied for other stars. We therefore choose to use the same LD coefficients for both quiet and active regions, but note this may not be fully accurate. We therefore include the option to use different sets of LD coefficients for active and quiet regions, should the user prefer to do so. 



\subsection{Stellar variability model}\label{section:2.4}

\begin{figure*}
\includegraphics[width=18cm]{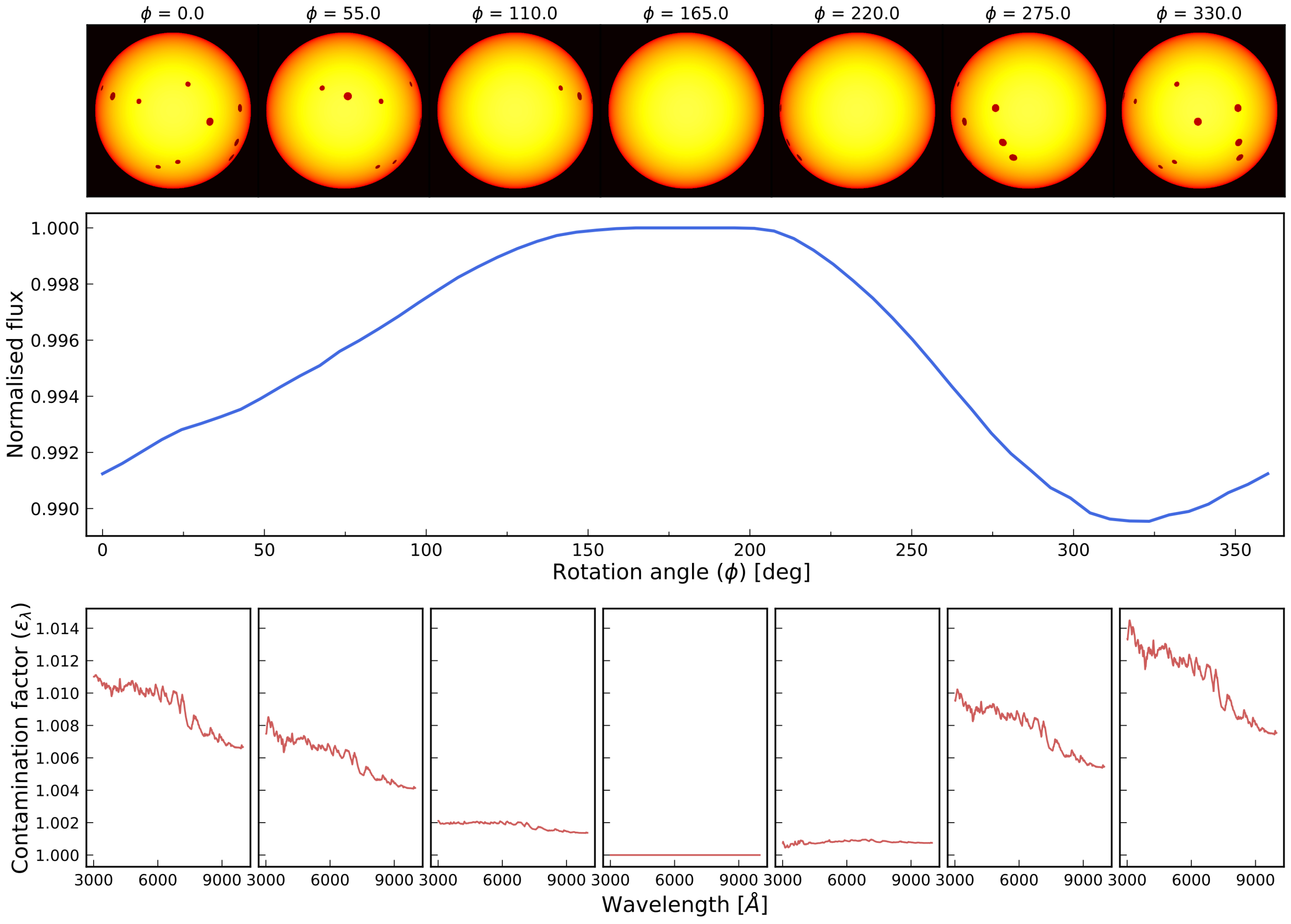}
  
  \caption{\textit{Top:} Stellar surface grid at different rotational phases for an (equator-on) K5V star with 10 2-3$^{\circ}$ spots randomly distributed across its surface. The effective temperature of the spots and the clear photosphere is 3500 K and 4500 K, respectively.  \textit{Middle:} The disc-integrated flux, calculated between 3000 and 10000 \r{A} as a function of the stellar rotational phase. \textit{Bottom:} The wavelength-dependent stellar contamination factor ($\epsilon_{\lambda}$) at rotation phases corresponding to those shown in the top panel.
     }
  \label{light curve rotations}
\end{figure*}


To compute the stellar flux variability for any specific configuration of active regions, we rotate the stellar grid model at the rotational period of the star. In addition, the user can set the stellar inclination ($i_{\star}$) i.e. the angle between the stellar rotation axis and the line of sight. We assume the star is a rigid rotator with non-evolving, circular active regions. The disc-integrated spectra are calculated at each rotational phase of the star and light curves are calculated by summing the emergent flux over the desired bandpass. The amplitude of the flux variations depends on the properties of the active regions and the bandpass of the instrument. For example, many Sun-like spots (of angular size $\sim$ 2$^{\circ}$) uniformly distributed across the photosphere will produce little variability. On the other hand, large M-dwarf type spots (of angular size of $\sim$ 7-8$^{\circ}$) covering the same total area as in the above case, will produce a larger amplitude, as they are less evenly distributed.

Figure \ref{light curve rotations} show the rotational variability model for a randomly chosen spot configuration on a K5V type, equator-on star (i$_{\star}$= 90$^{\circ}$). The light curve is calculated assuming a uniform throughput over a wavelength range of 3000-10000 \r{A}. 

One of the main limitations when inferring the properties of the stellar surface using the flux variability model is the degeneracy between the different properties of the active regions. The location, number, temperature, and size of the active regions are strongly correlated with each other. This can be remedied to some extent with constraints on the properties of the active regions obtained from other independent techniques. The occultation of an active region by a transiting planet results in the characteristic "bumps" in the transit light curve, which has been proven to be useful in measuring the temperature and size of the active region \citep[e.g.,][]{2011MNRAS.416.1443S}. While adding important information, spot crossings have only been reported for a handful of active stars \citep{2010A&A...510A..25S, 2011ApJS..197...14D, 2016AJ....151..150M, 2017ApJ...846...99M}. Another technique involves simultaneous photometric observations of the star in different filters. The amplitude of the flux variability is wavelength-dependent and by comparing the amplitude in different filters, the temperature of the active regions can be constrained \citep{1995ApJS...97..513H, 2020MNRAS.497.4602F}.

\subsection{Stellar contamination of transmission spectra}\label{Subsection: 2.5}


%

%

At any given time, the effect of unocculted active regions on the observed transit depth at different wavelengths can be described as follows:
\begin{equation}
    \left( \dfrac{R_{p}}{R_{\star}} \right)^{2}_{\lambda, \textrm{observed}} = \dfrac{\left( \dfrac{R_{p}}{R_{\star}} \right)^{2}_{\lambda, \textrm{true}}}{ 1 - f_{\textrm{proj}} \left( 1 - \frac{F_{\textrm{active}, \lambda}}{F_{\textrm{clear}, \lambda}} \right)} \quad\cdot
\end{equation}
Here, R$_{\rm{p}}$ is the radius of the planet, R$_{\star}$ is the radius of the star, $f_{\textrm{proj}}$ is the total projected surface area of inhomogeneities and $F_{\textrm{active}, \lambda}$ and $F_{\textrm{clear}, \lambda}$ are the spectrum of the active and clear photosphere, respectively (see also Equation (7) of \citealp{2011ApJ...736...12B}, Equation (1) of \citealp{2017ApJ...844...27Z} and Equation (1) of \citealp{2018ApJ...853..122R}). Thus, the measured and true transit depth are related by a wavelength-dependent factor ($\epsilon_{\lambda}$),
\begin{equation}{\label{Rackham conta}}
    \epsilon_{\lambda, t} = \frac{1}{1 - f_{\textrm{proj}}(t) \left( 1 - \frac{F_{\textrm{active}, \lambda}}{F_{\textrm{clear} \lambda}} \right)} \quad\cdot
\end{equation}
$\epsilon_{\lambda, t}$ is often referred to as the stellar contamination factor \citep{2018ApJ...853..122R}, the stellar activity correction factor ($\zeta$) \citep{2017ApJ...844...27Z}, or the contamination spectrum.
The relative change in observed transit depth at any specific wavelength is given by:
\begin{equation}{\label{transit depth}}
    \delta_\mathrm{dF} = \epsilon_{\lambda, t} - 1
\end{equation}
In \texttt{SAGE}, we calculate $\epsilon_{\lambda, t}$ as the ratio between the disc-integrated spectrum of the clear and the spotted star at the time of interest i.e.

\begin{equation}{\label{SAGE conta}}
\begin{split}
    \epsilon_{\lambda, t} & = \frac{\textrm{clear stellar spectrum} }{ \textrm{spotted stellar spectrum}} \\
    & = \frac{\sum_{j}^{N} F_{\textrm{clear}},_{\lambda},_{j}}{ \sum_{i}^{N_{\textrm{active}}} F_{\textrm{active}},_{\lambda},_{i} +  \sum_{j}^{N - N_{\textrm{active}}} F_{\textrm{clear}},_{\lambda},_{j} } \quad\cdot
\end{split}
\end{equation}


\noindent This method allows to accurately include projection effects of active regions and any additional atmospheric effects like LD. When the above secondary effects are neglected, Equation \ref{SAGE conta} transforms into Equation \ref{Rackham conta}  (see Appendix \ref{Section: 7.1}).

\begin{figure*}
\centering
    \includegraphics[width=18cm]{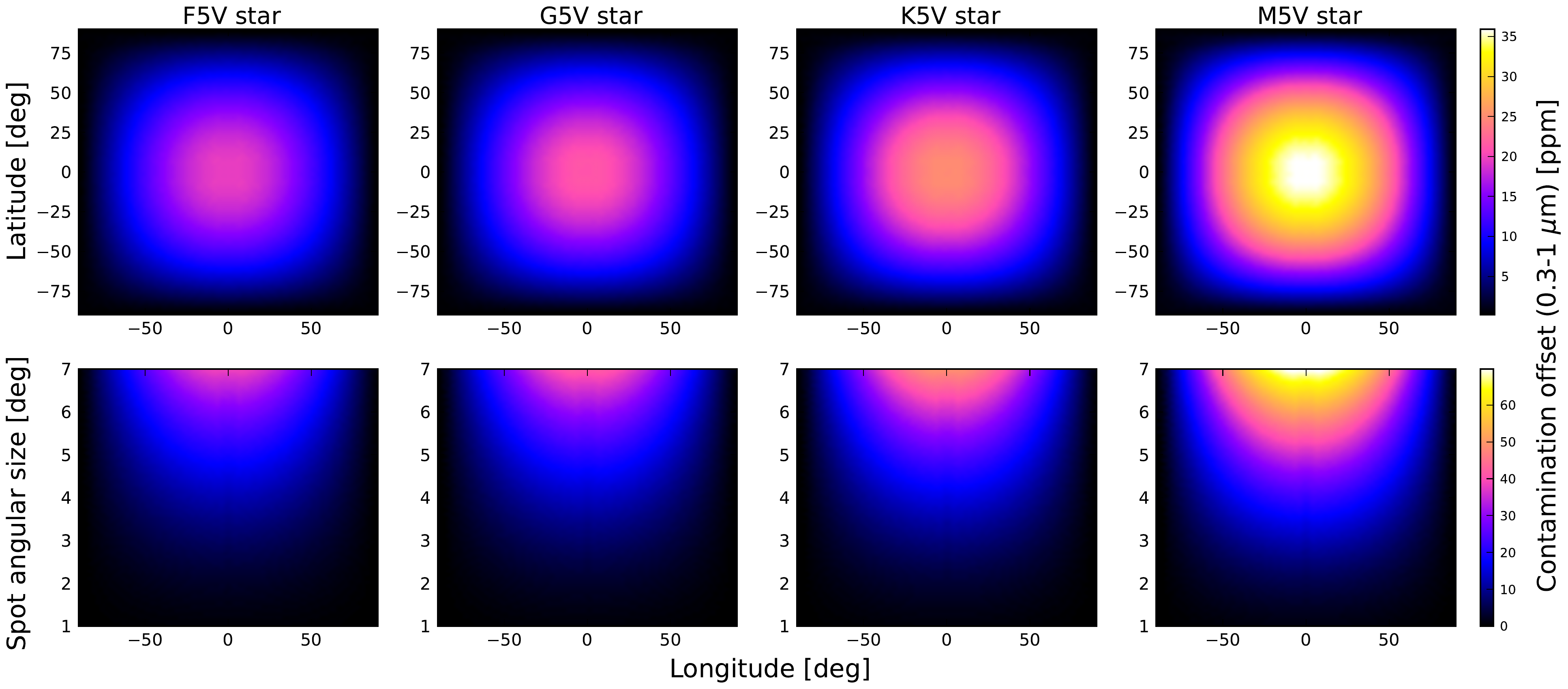}
     \caption{\label{fig:noLD}\textit{Top:} Stellar contamination offset for a single spot of 5$^{\circ}$ angular size, placed at different locations on the  stellar disc for stars of different spectral types. \textit{Bottom:} Stellar contamination offset for the same stars with spot sizes ranging from 1 to 7$^{\circ}$ for different longitudes. LD was not included in these simulations. 
     }
     \label{ Figure: 5}
\end{figure*}

\section{The effects of starspots on transmission spectra}\label{Section 3}

In the previous section, we presented \texttt{SAGE} and the different stellar effects which can be modelled via the pixelation approach. Here, we investigate the impact of projection and limb-darkening effects on the extent of stellar contamination in the transmission spectrum. We here only consider dark starspots as the limb-darkening behaviour of faculae is little understood at this time. We study stellar contamination for F5V, G5V, K5V and M5V type stars, using model input spectra from the PHOENIX spectral database \citep{2013A&A...553A...6H}. Where necessary, spectra at specific temperatures are calculated by linear interpolation of neighbouring grid points. Table \ref{table:1} contains the adopted stellar parameters for both clear and spotted regions for different spectral types. The temperatures of the clear photospheres are obtained from \cite{2013ApJS..208....9P} and the spot temperatures are calculated from \cite{2018ApJ...853..122R, 2019AJ....157...96R}, which uses the relation T$_{\rm{spot}}$= 0.418 $\times$ T$_{\rm{clear}}$ + 1620 K for F, G and K type stars and T$_{\rm{spot}}$= 0.86 $\times$ T$_{\rm{clear}}$ for M-dwarfs. The surface gravity (log g) is set to 4.5 for F, G and K stars and to 5.5 for the M-dwarf. In all stars, we assume solar metallicity with no alpha enrichment.

\subsection{Position and size of active regions}\label{Section: 3.1}

The extent of stellar contamination on the transmission spectra depends mainly on two properties of the active regions: their projected surface area (a position and size-dependent property) and the surface brightness contrast with respect to the surrounding photosphere. The projected surface area of an active region changes not only with stellar rotation but also due to their evolution over time. Note that stellar rotation alone has no impact on the absolute filling factor of the active regions\footnote{As defined in Sec.\,\ref{Section 2.1}, the absolute filling factor refers to the ratio of the \textit{total} surface area of the active regions to the surface area of the star, while the projected filling factor is the ratio of the \textit{projected} surface area of active regions to the surface area of the stellar disc.}. 

We studied the dependence of stellar contamination on the projected surface area of spots through simulations with \texttt{SAGE}, assuming spot sizes between 0 and 7$^{\circ}$ and locations ranging from the center of the stellar disc to the limb of the star. For this, we assumed a uniformly-illuminated stellar disc with no LD and rotational broadening. The simulations were performed for all four spectral types tabulated in Table \ref{table:1}.

As is visible in Fig.\,\ref{light curve rotations} (bottom panel), the stellar contamination rises towards bluer wavelengths. To quantify the strength of stellar contamination for different cases, we choose to use the change in observed transit depth at different wavelength bands, hereafter, the contamination offset. We define the contamination offset ($\delta$) as the difference between the average observed transit depth in between wavelength bands 3000 - 4000\,\r{A} ($\epsilon_{0.3-0.4}$)  and 9000 - 10000\r{A} ($\epsilon_{0.9-1.0}$ ).



\begin{equation}\label{contamination_offset_equation}
    \delta = \left(\frac{R_{p}}{R_{\star}} \right)_{\rm{true}}^{2} \times \left(\epsilon_{0.3-0.4} - \epsilon_{0.9-1.0}\right) \quad\cdot
\end{equation}
Here, $(R_{p}/ R_{\star})_{\rm{true}}^{2}$ is the true transit depth.  


In Fig. \ref{ Figure: 5}, we present the computed contamination offsets for different stars transited by a hypothetical, bare-rock, planet i.e. a planet with no atmosphere and a uniform transit depth of 1$\%$ in all wavelengths. In the top panels, we illustrate the offsets calculated for spots located at different locations on the stellar disc. The dependence of stellar contamination on the position of spots is clearly visible. In all simulations, the effect is maximal when the spot is located at the center of the stellar disc, but decreases significantly when the spot is moved closer to the limb. For the M5V-type star, the maximum offset exceeds 30\,ppm. However, we do not measure such high offsets for the earlier spectral types, even though they have cooler spots. Generally, we find a clear anti-correlation between the contamination offset and the effective temperature of the star with the cooler stars having a higher contamination offset than the hotter stars.

%

\setlength{\tabcolsep}{10pt}
\renewcommand{\arraystretch}{1.4}
\begin{table}
\caption{Model spectrum parameters used in the presented simulations}             
\label{table:1}      
\centering                          
\begin{tabular}{c c c c}        
\hline\hline                 
Spectral type & T$_{\rm{clear}}$ [K] & T$_{\rm{spot}}$ [K] & log g [cm sec$^{-2}$] \\    
\hline                        
   F5V & 6510 & 4341 & 4.5 \\      
   G5V & 5660 & 3986 & 4.5 \\
   K5V & 4450 & 3480 & 4.5 \\
   M5V & 2880 & 2477 & 5.5 \\
\hline                                   
\end{tabular}
\end{table}

The bottom panels of Fig. \ref{ Figure: 5} illustrate the dependence of the contamination offset on the spot size. As expected, contamination offsets are largest for large spots located close to the center of the stellar disc. Again, M5V stars show the highest contamination, with offsets exceeding 60\,ppm for spots with an angular size of 7$^{\circ}$. Such large spots have been detected in M-dwarfs through occultations by a transiting planet \citep{2022A&A...667L..11A}. These simulations isolate the dependence of the variations of the stellar contamination with the projected surface of the active regions, as we assumed a non-limb darkened star emitting equally at all distances from the center. 

\begin{figure*}
\centering
    \includegraphics[ angle=-90, width=18cm]{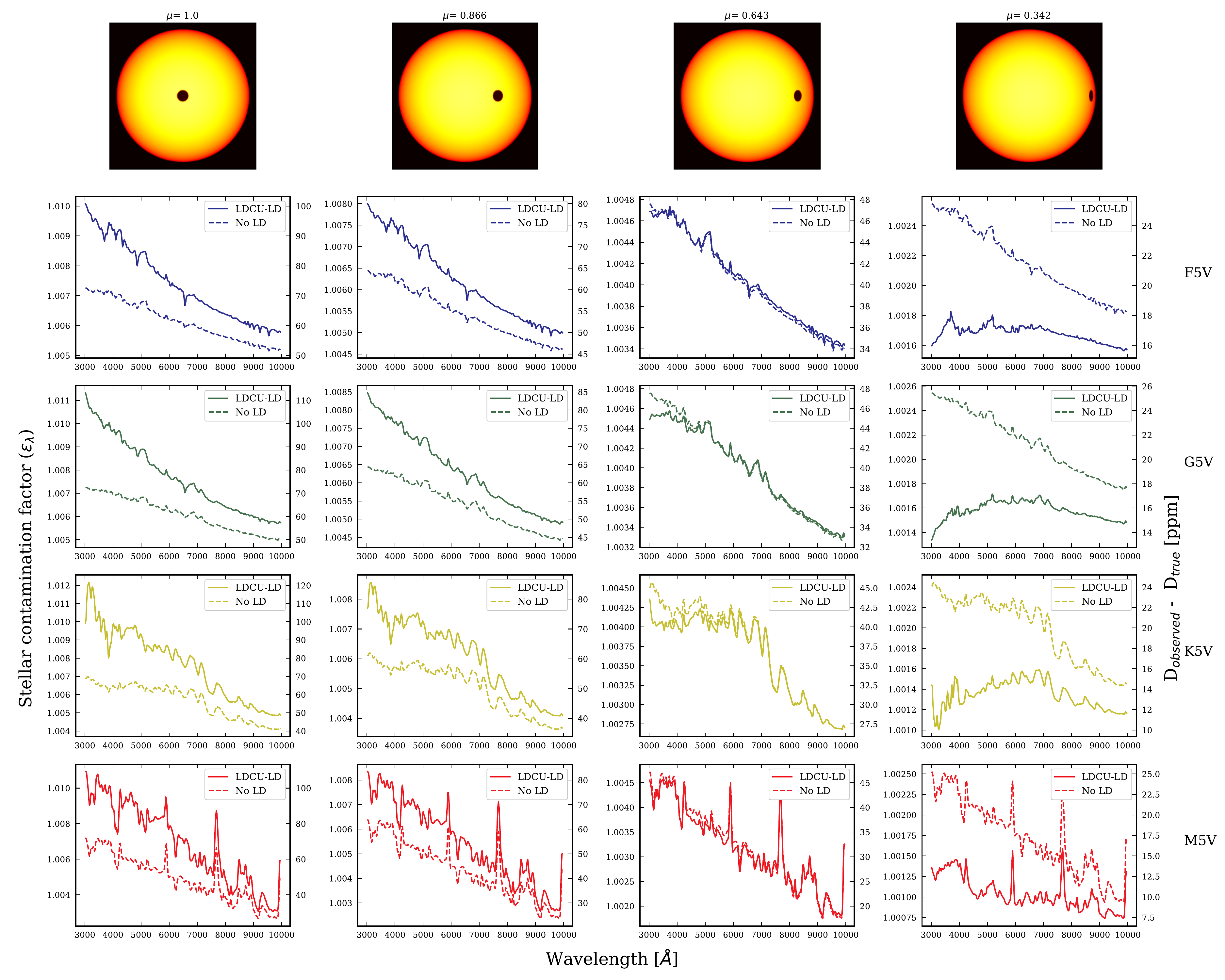}
     \caption{\textit{Top:} Illustration of the stellar disc with a single spot at 4 different positions, i.e., at different stellar rotation phases. \textit{Bottom:} The corresponding stellar contamination spectra for F5V, G5V, K5V and M5V stars with (solid) and without (dashed) LD. The secondary y-axis represents the change in transit depth due to stellar contamination assuming a planet with a uniform transit depth of 1\% at all wavelengths.
    }
     \label{Figure: 6}
\end{figure*}

\subsection{Limb-darkening}{\label{Section 4.2}}

A more physically accurate simulation includes the effect of LD, i.e., the drop in intensity between the center of the stellar disc and its limb. We thus recreated our simulations incorporating LD in the pixelated stellar grid. To do so, we adjusted the intensity of every pixel on the stellar grid following a quadratic LD law. The coefficients are calculated at each wavelength using \texttt{LDCU} with the stellar parameters T$_{\rm{eff}}$ and log(g) as tabulated in Table\,\ref{table:1}.



\begin{figure*}
\centering
    \includegraphics[width=18cm]{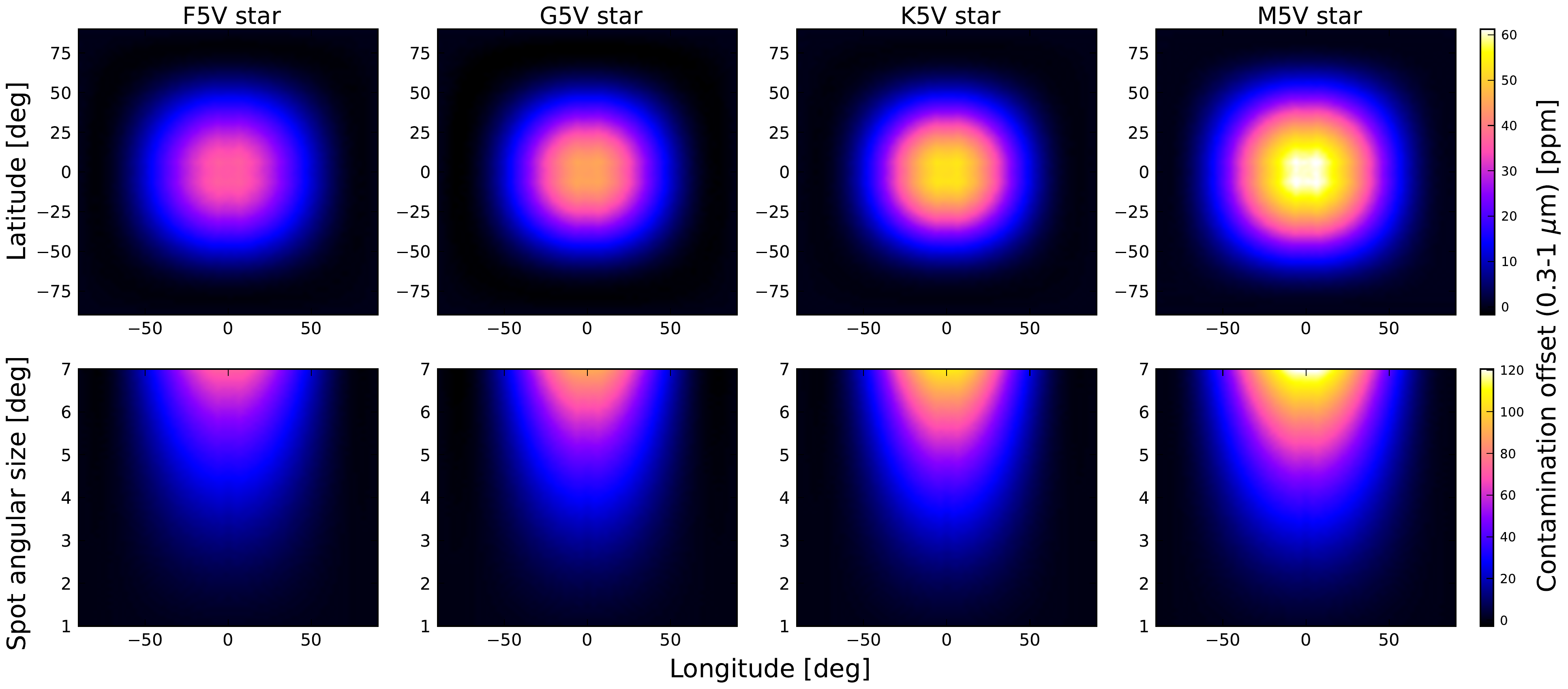}
    \caption{As Fig. \ref{fig:noLD}, but including LD.}
     \label{Figure: 7}
\end{figure*}

In Fig.\,\ref{Figure: 6}, we illustrate the effect of LD on the stellar contamination spectra using as an example a 5$^{\circ}$ spot at longitudes of 0$^{\circ}$, 30$^{\circ}$, 50$^{\circ}$ and 70$^{\circ}$ for the four spectral types considered. The contamination spectra show a diversity of slopes and features for different spectral types and spot positions. For all spectral types, the LD model predicts higher contamination than the non-LD case when the spot is located closer to the center of the stellar disc ($\mu= 1$ and $\mu = 0.866$). Indeed, the more realistic LD simulations show a $\sim$1.8 $\times$ larger contamination offset (calculated using Eq. \ref{contamination_offset_equation}) , compared to the simplified non-LD case. On the other hand, in the case of spots being located closer to the limb of the star, the non-LD model overestimates the contamination effect. Here, the offset is up to $\sim$0.5 $\times$ reduced compared to the non-LD model. This behaviour is easy to understand: due to limb-darkening, the bright disc center contributes more strongly to the total flux, enhancing the impact of spots located there. Conversely, the impact of spots near the darker limb is reduced.

LD also has an impact on the shape of the contamination spectrum due to its wavelength dependence. For spots close to the disc center, the characteristic rise in the contamination spectrum at lower wavelengths is visible in both the LD model and the non-LD model. However, for spots located very close to the limb of the star, LD considerably flattens the contamination spectra, most notably in the wavelength regime of 4500 to 7500 \r{A}. In addition, at wavelengths $< 4500 $ \r{A}, spots at the stellar limb produce a positive slope in the spectra for the F5V, G5V and K5V-type stars. 

We can again compute the contamination offset $\delta$ (similar to Sec.\,\ref{Section: 3.1}) and study its variation with the position and size of the active regions. The results are summarised in Fig.\,\ref{Figure: 7}. For spots located closer to the center of the stellar disc, the amplitude of the stellar contamination offset increases by roughly 30 ppm with the addition of limb darkening. For the M-type star, spots with an angular size of 7$^{\circ}$ can introduce transit depth offsets exceeding 120 ppm. Contrary, the contamination offset reduces strongly for spots closer to the limb with the addition of LD.


\section{Stellar surface retrievals} \label{Section 4}

\begin{figure*}
\centering
    \includegraphics[width=18cm]{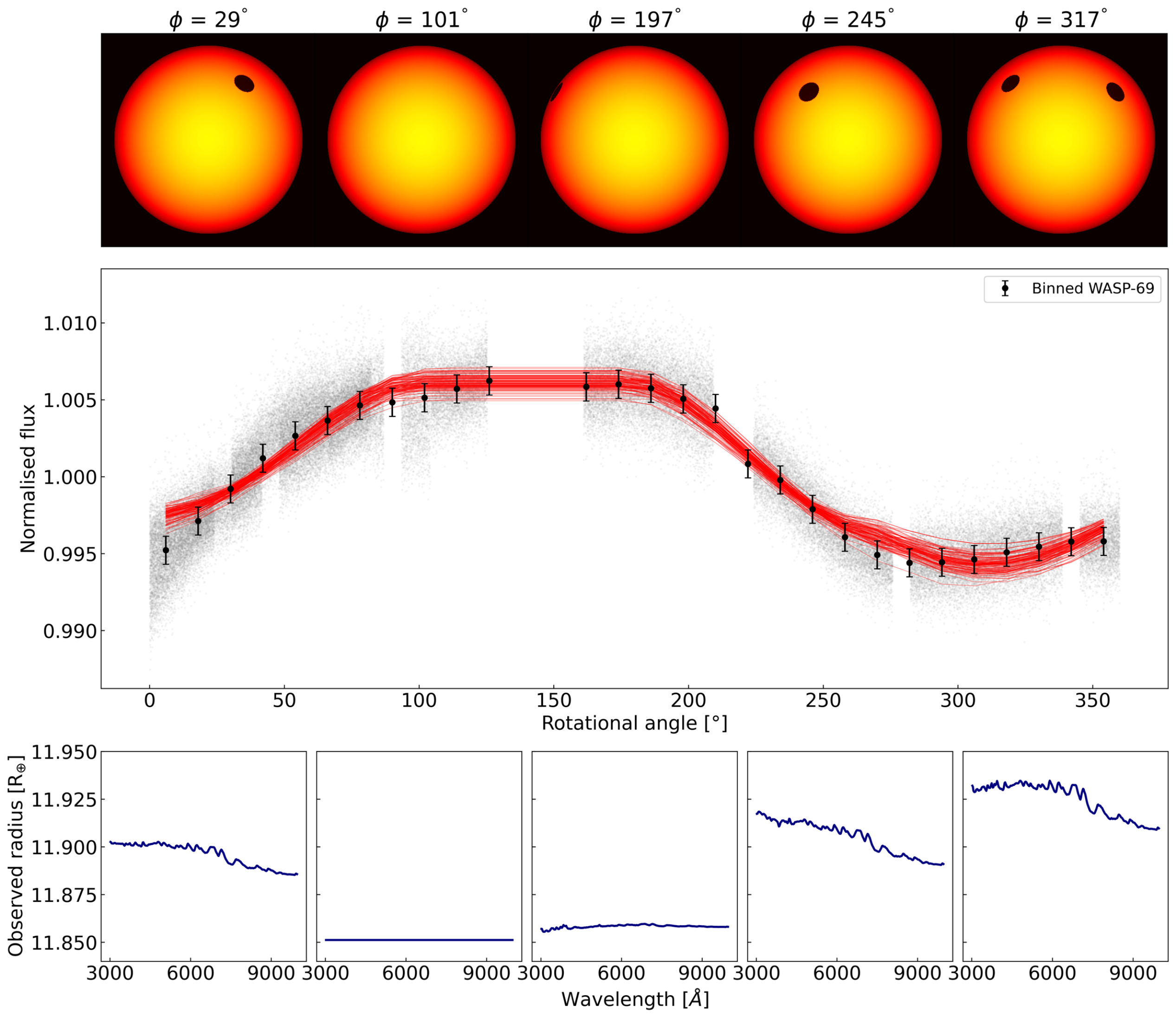}

    \caption{\emph{Top:} A model stellar surface map for variability model associated with the peak of the posteriors. \emph{Middle:} TESS data of WASP-69, phase-folded on the stellar rotation period. The grey points are the individual measurements, and the black points are the binned measurements. The red curves show 100 variability models randomly selected from the posterior distribution of the fit. \emph{Bottom:} 
    The observed radius of planet b at different rotational phases of the star and different wavelengths.}
    
     \label{Wasp-69 variability model}
\end{figure*}

\subsection{Model setup}\label{section:4.1}

\texttt{SAGE} can be used to retrieve the properties and positions of active regions from observations of the rotationally-modulated flux of the star. To do so, we use the capacity of \texttt{SAGE} to predict the photometric variability produced by a given spot configuration by rotating a stellar surface model as described in Sec.\,\ref{section:2.4}. Here, \texttt{SAGE} includes the full stellar atmospheric effects described above, most importantly the LD. Furthermore, the inclination of the star and its impact on the flux variability is also accounted for.


To derive the properties of the stellar surface inhomogeneities, we create model predictions from \texttt{SAGE} and use an MCMC framework to evaluate these against the observed data. The fitted parameters are: 
\begin{itemize}
    \item The stellar rotation period
    \item The stellar inclination, i.e., the angle between the sky plane and the stellar rotation axis
    \item For each active region: latitude, longitude, size and contrast
    \item A jitter term to account for any additional photometric uncertainties
\end{itemize} 
We recommend imposing uniform distributions as priors for all parameters, unless prior information is available about the distribution of active regions. The width of the priors (for latitude and longitude) can be set to cover the entire stellar surface or restricted to remove the possibility of polar spots.

One of the main limitations in accurately constraining stellar surfaces using the variability models is that the retrievals are performed under the assumption that the observed flux variability of a star is due to a limited number of large active regions. Thus, this technique is unable to capture the variability associated with many small inhomogeneities, randomly distributed across the stellar surface. However, it does provide accurate estimates of the \textit{relative} coverage of the active regions over the different rotation phases of the star. 

\subsection{Application to WASP-69}\label{section:4.2}

We used the setup described above to study the flux variability of the exoplanet host WASP-69 in data collected by the Transiting Exoplanet Survey Satellite (TESS; \cite{2014SPIE.9143E..20R}). The K5\,-type star has an effective temperature of 4715 $\pm$ 50 K and hosts a puffy Saturn-mass planet (WASP-69b, 0.26M$_{\rm{J}}$, 1.06R$_{\rm{J}}$), orbiting with a period of 3.87d \citep{2014MNRAS.445.1114A}.

\citet{2014MNRAS.445.1114A} detected emission features in the Ca\,H + K line cores, indicative of activity, and estimated an activity index of $\mathrm{log\, R^{'}_{HK} \approx -4.54}$, which puts the age of the star at $\sim$ 0.8 Gyr \citep{2008ApJ...687.1264M}. The stellar rotational period calculated from light curve modulation is 23.07 $\pm$ 0.16 d, broadly agreeing with that derived from the spectroscopic rotational broadening method, $<$ 18.7 $\pm$ 3.5 d. This suggests that the star is seen equator-on. 

The atmosphere of the planet WASP-69b has been studied using the transmission spectroscopy technique in both low and high resolution, which has led to robust detection of multiple chemical species such as CH$_{4}$, NH$_{3}$, CO, C$_{2}$H$_{2}$ and H$_{2}$O \citep{2022A&A...665A.104G}. The presence of Rayleigh scattering  by molecular hydrogen in the upper atmosphere has also been detected \citep{2020A&A...641A.158M}. However, unocculted spots on the stellar surface are speculated to be a possible alternative explanation for the bluewards rise in transit depth. 

\setlength{\tabcolsep}{10pt}
\renewcommand{\arraystretch}{1.4}
\begin{table}
    \centering
    \begin{tabular}{c c c}
       \hline
       \hline
       Jump parameter & Prior & Results \\
       
       \hline
       Spot - 1 & & \\
       Latitude [deg]  &  $\mathcal{U}$(90, 0)  & 41.74$^{+16.999}_{-26.753}$ \\
       
       Longitude [deg]  & $\mathcal{U}$(-180, 180) & -0.72$^{+4.265}_{-4.713}$ \\
       
       Size [deg]      & $\mathcal{U}$(1, 10) & 6.91$^{+2.153}_{-1.189}$\\
       
        & & \\
        Spot - 2 & & \\
       Latitude [deg]   & $\mathcal{U}$(90, 0) & 31.23$^{+20.210}_{-19.663}$ \\
       
       Longitude  [deg] & $\mathcal{U}$(-180, 180) & 85.22$^{+3.679}_{-4.001}$ \\
       
       Size  [deg]    & $\mathcal{U}$(1, 10) &  6.72$^{+1.830}_{-0.651}$\\
       \hline
    \end{tabular}
    \caption{Fitted spot parameters and their priors for the stellar surface retrievals of WASP-69. Here, $\mathcal{U}$(a,b) specifies a uniform prior between a and b. }
    \label{tab:results-surface-retrivals}
\end{table}

Here, we use the recently acquired observations of WASP-69 from TESS Sector 55 (UT 2022-August-05 to UT 2022-September-01). We used the 20\,s cadence light curves generated by the Science Processing and Operations Center (SPOC; \citealp{2016SPIE.9913E..3EJ}) and employed the \texttt{lightkurve} package \citep{2018ascl.soft12013L} to calculate the flux using Simple Aperture Photometry (SAP).  The flux outliers are removed using a running median method. The points above 3$\times$MAD\footnote{Mean Absolute Deviation from the median for each 15-point window} are removed along with the 6 transits of WASP-69b. In Sector 55, the median background flux increased by a factor of 4 between 2805 and 2807\, TESS Barycentric Julian Date (TBJD) due to stray light from the Moon, and we removed all observations taken during this time window. A clear offset is present in the light curve at 2806\,TBJD. We calculate the average flux over a 2-day window before and after the offset and shifted the data taken before 2806\,TBJD accordingly to remove this offset. The processed light curve shows a periodicity in flux with a period of  $\sim$ 22.1\,d, in good agreement with the previously-determined stellar rotation period. As the total science data is collected over a period of 26.26 days, we obtain a little over one complete rotation of the star. In addition, we found no evidence of stellar flares in the data. For the sake of computational speed, we phase-folded the data (with P$_{\rm{rot}}$ = 22.1 d) and binned them into 28 uniformly-spaced time intervals before progressing with the stellar surface retrievals. 

To limit the degeneracy between temperature and size of active regions, we fixed the contrast of the spots in our simulations. We set the clear photospheric temperature to 4715\,K and, using the linear relation between the clear and spotted photosphere discussed in Sec. \ref{Section 3}, we set the temperature of spots to 3590\,K. We convolved the PHOENIX spectra of both clear and spotted photospheres with the TESS bandpass, finding a spot contrast ($\rho_{\rm{spot}}$) of 0.26. We used LDCU to compute quadratic LD coefficients assuming the temperature of the clear photosphere, and we kept these coefficients fixed during the retrieval. The system is likely seen equator-on, evidenced by the fact that the stellar rotation period computed from the $V$ sin($i_{\star}$) (< 18.7 $\pm$ 3.5 d) and the rotation period derived from the TESS light curve (22.1 d) agree. We, therefore, fix the stellar inclination to 90$^{\circ}$.


We assumed uniform priors for latitude, longitude and size of each spot, restricting the latitudes to the northern hemisphere of the star to avoid bimodal posteriors due to the symmetry between North and South. The spot size was limited to be between 1$^\circ$ and 10$^{\circ}$. This range covers small Sun-like spots to large M-dwarf-type spots. To determine the number of spots, we first ran a retrieval with a single spot and then iteratively increased the number of spots from 1 to 5. Each time, we also calculated the absolute filling factor of the active regions. The  2-spot model has the lowest Bayesian Information Criterion (BIC), and we thus deem it the most appropriate. We also find that increasing the number of spots to model the light curve does not have a significant impact on the absolute filling factor of the active regions. 

In Fig.\,\ref{Wasp-69 variability model}, we present the phase-folded TESS data together with 100 variability models randomly selected from the posteriors of our MCMC analysis. We also show a model of the stellar surface corresponding to the peak of the posterior distribution. We present the posteriors of our retrieval in Fig. \ref{corner-surface-retrivals} and list the derived parameters together with their uncertainties in Table \ref{tab:results-surface-retrivals}. While a range of spot latitudes is possible given the available data, our analysis strongly rejects the possibility of polar spots on WASP-69. From the posterior distribution, the correlation between the spot size and latitude is clearly visible; the spots at higher latitudes are larger and vice-versa. This relation preserves the projected filling factor of the active regions which modulate both the stellar flux and contamination.

We measure the absolute filling factor of the active regions to be f$_{s, \textrm{absolute}}$ = 0.77 $^{+0.29}_{-0.19}$ \%. Assuming a uniform radius of planet b ($R_{p}$= 11.85$R_{\oplus}$) at different wavelengths, we also constrained the stellar contamination in the optical regime and its variation with the different rotation phases of the star as shown in Fig. \ref{Wasp-69 variability model}.

To probe the impact of the i$_{\star}$=90$^{\circ}$ assumption, we also performed a separate analysis with Gaussian priors on the inclination ($\mu$=90$^{\circ}$, $\sigma$=15$^{\circ}$), truncated at < 90$^{\circ}$ to avoid degeneracies due to the symmetry of the problem. Here, the $\sigma$ of 15$^{\circ}$ is calculated by taking the lower limit on inclination derived using $V$ $sin(i)$ of 2.2 $\pm$ 0.4 km sec$^{-1}$, a stellar radius of 0.813 $\pm$ 0.028 $R_{\odot}$ and P$_{\rm{rot}}$  of 22.1 days \citep{2014MNRAS.445.1114A}. We used uniform priors on the entire stellar surface and spot sizes between 1$^{\circ}$ to 10$^{\circ}$. The posteriors of the analysis are presented in Fig. \ref{corner-surface-retrivals_with_inclination}. The retrievals qualitatively reproduce those found for i$_{\star}$ set to 90$^{\circ}$, namely two spots at mid-latitudes separated by 85$^{\circ}$ in longitude. The posterior for the inclination peaks at 68$^{\circ}$, but extends out to include solutions with 90$^{\circ}$. Due to the inclination the symmetry between north and south is broken, and the derived spot sizes depend on the hemisphere on which the spot is located. While we adopt the 90$^{\circ}$ inclination for WASP-69 due to the limited quality and quantity of data available, this assumption might need revision once more detailed observations are available.

\section{Discussion}

In this paper, we have used a stellar surface pixelation approach to investigate the two-fold impact of active regions on exoplanet observations. First (see Section\,\ref{Section 3}), we studied their impact on transmission spectra, including a thorough treatment of projection effects and LD. Second (see Section\,\ref{Section 4}), we illustrated the photometric effect of spots by rotating our model and devised a retrieval approach to constrain the properties of active regions from photometric observations. 

\subsection{Advantages and novelty}

For all but the most quiet stars, the stellar surface, and most importantly the quantity and position of active regions vary with time, leading to varying stellar contamination in transmission spectra. This makes it difficult to combine multi-epoch transit observations to enhance detection of atmospheric features. One of the main advantages of \texttt{SAGE} is its ability to use the stellar surface map, obtained using  photometric time series observations, to directly constrain the stellar contamination at different rotation phases of the star. This feature is currently not included in tools like \texttt{Starry} \citep{2019AJ....157...64L} and \texttt{SOAP} \citep{2012A&A...545A.109B}, which are mainly used to create stellar surface maps from photometric observations. In addition, \texttt{SAGE} also calculates the stellar contamination models using a numerical approach accounting for LD, which is more precise than the purely analytical models. 

An important result of our study is the impact of limb darkening on stellar contamination. As an active region moves closer to the limb of the star, the overall surface brightness of the stellar photosphere decreases, and so does its impact on the measurements. In the case of spots located close to the center of the stellar disc, the non-LD model \textit{underestimates} the stellar contamination. In contrast, for spots closer to the limb, the non-LD model \textit{overestimates} the stellar contamination. This result has been tested with multiple simulated stars of different spectral types with different spot temperatures. 
 
As LD is a chromatic effect, the changing surface brightness of an active region as it approaches the limb also has a wavelength-dependent effect on the observed transit depths. The LD effect is stronger at shorter wavelengths, impacting the shape of the contamination spectrum. In Fig. \ref{Figure: 6}, we presented the diversity in contamination spectra for F5V, G5V, K5V and M5V stars with spots at different distances from the center of the stellar disc. In all cases, as a spot moves close to the limb of the star, the slope of the contamination spectrum flattens significantly for the wavelength range 4500 - 7500\r{A}. At shorter wavelengths of 3000 - 4500 \r{A}, the slope can even invert for F, G and K-type stars, leading to a decrease in contamination at short wavelengths. This is contrary to non-LD models, which predict a monotonous rise in stellar contamination towards shorter wavelengths, an effect that can be misinterpreted as atmospheric Rayleigh scattering in exoplanets \citep{ 2014A&A...568A..99O,2020A&A...641A.158M}. Thus, the inclusion of LD in stellar contamination calculations is vital for disentangling planetary and stellar effects on the transmission spectrum.

\subsection{Further development and caveats}

Although \texttt{SAGE} takes into account more stellar atmospheric effects compared to the existing routines, it is still limited by the complexity of stellar atmospheres, and we have made some simplifying assumptions in this study.

One of the central assumptions of the model is that the star is rotating as a rigid body. This is not the case, as observed clearly for the Sun, which is known to show differential rotation \citep{1986A&A...155...87B,2000SoPh..191...47B}, rotating significantly faster at the equator than at the poles. Not including for differential rotation can affect our retrievals of stellar surface properties from photometric flux variability, as the contributions of equatorial and polar spots will have different timescales. Similarly, we also assume that active regions are not evolving within the investigated timescales. This assumption works well for fast rotators where the stellar rotation period is much smaller than the evolutionary timescales of active regions such as AU Mic \citep{2021A&A...654A.159S}, or for stars with long-lived and stable spots that last a few stellar rotations, e.g. HD 189733 and $\alpha$ Cen B \citep{2014ApJ...796..132D}.

Stellar surface maps are also affected by an important degeneracy between the position, size and temperature of active regions. The flux variability due to a hot, large spot (or a cluster of small spots) on the stellar surface can be reproduced equally well with a smaller, cooler spot. The potential presence of multiple stellar spots with different temperatures and sizes further complicates the scenario. To break this degeneracy, an independent measurement of the size or temperature of the active regions is needed. This may be attempted with multi-colour observations, or transit observations that show short-term brightening due to the starspot crossings. Another complexity lies in the LD behaviour of active regions. While we assume the same LD coefficients for active regions as for the clear photosphere, their LD behaviour might differ in reality. 

Finally, we currently do not include high-resolution effects such as the convective blueshift and variations in line shapes, limiting the applicability of the code to low and medium-resolution observations. While updates are foreseen in future releases, we expect these effects to be minimal for the results presented in this paper. 

Using the pre-tabulated models of stellar emission spectra, we also assume that the emergent flux from each photospheric pixel is the same as that of a disc-integrated spectrum of a star with the same temperature, metallicity and gravity. This assumption is commonly used in this field \citep{2008MNRAS.385..109P, 2014ApJ...791...55M, 2017ApJ...834..151R}, but we do acknowledge the need for more sophisticated spectral models. 

 
\section{Summary \& Conclusions}

In this paper, we discuss the effect of stellar activity on exoplanet transmission spectra using realistic models of stellar photospheres implemented through a pixelation approach. In particular, we investigate the effect of the spot position and stellar limb darkening on the contamination spectrum at optical wavelengths (3000-10000\r{A}). The developed tool, \texttt{SAGE}, not only provides contamination spectra needed to correct exoplanet transmission spectra, but also allows to derive stellar surface maps from photometric stellar rotation curves. Our key findings are summarised below:

\begin{enumerate}

\item The extent of stellar contamination depends on the effective temperature of the star. From simulations of F5V, G5V, K5V and M5V stars, we detect a clear negative correlation between the temperature of the star and the amplitude of the stellar contamination effect. 

\item The shape and amplitude of stellar contamination depends strongly on the location of the spot. Due to projection effects, spots near the centre produce more pronounced effects. 

\item Stellar limb-darkening further exacerbates this projection effect. For spots located close to the center of the stellar disc ($\mu$ $\gtrsim$ 0.6), the stellar contamination effect is enhanced due to limb-darkening, but for spots located close to the limb ($\mu$ $\lesssim$ 0.6), the contamination decreases.

\item  Limb-darkening also affects the shape of the contamination spectrum due to its colour dependence, especially for active regions located close to the limb of the star. For spots near the limb, we find a flattened contamination spectrum between 4500 - 7500\r{A} and even a negative slope for < 4500 \r{A} for the simulated F, G and K-type stars. Thus, the monotonous rise in transit depth associated with unocculted spots depends on the location of spots. 

\item We also used \texttt{SAGE} to model the flux variability of WASP-69 in TESS observations. We find that the $\sim$ 22.1 d periodicity in the flux can be modelled as an effect of active regions on the stellar surface. We estimate the spots to cover $\sim$ 1 \% of the total stellar surface. 
 \end{enumerate}
 
To conclude, the purpose of the developed tool is to calculate precise stellar contamination spectra, accounting for realistic and time-varying spot configurations, and coupling these computations with the photometric flux variability that encodes the distribution of spots on the star. The increasingly detailed understanding of stellar effects on transmission spectra will enable the meaningful comparison of multi-epoch transmission spectra, opening the door towards studies of variability in exoplanet atmospheres.

\begin{acknowledgements} HC acknowledges the fruitful discussions with Xavier Dumusque, Yinan Zhao, Alex Pietrow and Matthew Battley. HC, ML, BA and DP further acknowledge the support of the Swiss National Science Foundation under grant number PCEFP2\_194576. This work has been carried out within the framework of the NCCR PlanetS supported by the Swiss National Science Foundation under grants 51NF40\_182901 and 51NF40\_205606. DP further acknowledges funding from the European Research Council (ERC) under the European Union’s Horizon 2020 research and innovation programme (project {\sc Four Aces}; grant agreement No 724427).  AD and DP acknowledge financial support from the Swiss National Science Foundation (SNSF) for project 200021\_200726. This paper includes data collected by the TESS mission. Funding for the TESS mission is provided by the NASA's Science Mission Directorate. The computations in this paper were performed at the University of Geneva on the "Yggdrasil" HPC cluster. 

\end{acknowledgements}


\bibliographystyle{aa} 
\bibliography{biblography.bib}

\begin{appendix}

\section{Equivalence of stellar contamination calculations in case of no secondary effects on the stellar disc} \label{Section: 7.1}

Under the assumption of a planet transiting a uniformly illuminated stellar disk, the measured transit depth is given by:

\begin{equation}
   \delta_{\rm{transit, \lambda}} = \left( \frac{R_{p}}{R_{\star}} \right)^{2}_{\lambda}= \frac{F_{\rm{out, \lambda}} - F_{\rm{in, \lambda}}}{ F_{\rm{out, \lambda}
   } } \quad\cdot
\end{equation}

Here, $\delta_{\rm{transit}, \lambda}$ is the flux drop associated with the planetary transit, F$_{\rm{out, \lambda}}$ and F$_{\rm{in, \lambda}}$ are the spectra of the star measured before and during the transit.

In the case of a planet transiting an active host star with $F_{\rm{Clear}, \lambda}$ is the spectrum of the clear photosphere and $F_{\rm{Active}, \lambda}$ is the spectrum of an active region. The out-of-transit spectrum of the star (F$_{\rm{out, \lambda}}$) under the assumption of a stellar disk with no limb-darkening effect is:

\begin{equation*}
    F_{\rm{out, \lambda, t}}= [1- f_{\rm{proj}}(t)] F_{\rm{clear}, \lambda} + f_{\rm{proj}}(t) F_{\rm{active}, \lambda} \quad\cdot
\end{equation*}

Here, $f_{proj}(t)$ is the projected covering fraction of the active regions with respect of an observer at any particular time. Assuming  a planet with no atmosphere and a true transit depth of $\left( \frac{R_{p}}{R_{\star}}\right)^2_{\lambda}$, the in-transit spectrum ($F_{in, \lambda}$) of the star is given by:

\begin{equation*}
    F_{\rm{in, \lambda, t}}= [1 - f_{\rm{proj}}(t) - \left( \frac{R_{p}}{R_{\star}}\right)^2_{\lambda} ] F_{\rm{clear}, \lambda} + f_{\rm{proj}}(t) F_{\rm{active}, \lambda} \quad\cdot
\end{equation*}

\noindent
Thus, the observed transit-depth $\left( \Hat{\frac{R_{p}}{R_{\star}}} \right)^{2}_{\lambda, t}$  is given by:

\begin{equation*}
\begin{split}
    1 - \left( \Hat{\frac{R_{p}}{R_{\star}}} \right)^{2}_{\lambda, t} & =  \frac{F_{\rm{in, \lambda, t}}}{ F_{\rm{out, \lambda, t}}} \\
    & = \frac{[1 - f_{\rm{proj}}(t) - \left( \frac{R_{p}}{R_{\star}}\right)^2_{\lambda} ] F_{\rm{clear}, \lambda} + f_{\rm{proj}}(t) F_{\rm{active,} \lambda}}{[1- f_{\rm{proj}}(t)] F_{\rm{clear}, \lambda} + f_{\rm{proj}}(t) F_{\rm{active}, \lambda}} \\
    & = \frac{ 1 - f_{\rm{proj}}(t) + f_{\rm{proj}}(t) \frac{F_{\rm{active}, \lambda}}{F_{\rm{clear,} \lambda}} - \left( \frac{R_{p}}{R_{\star}}\right)^2_{\lambda}  }{ 1 - f_{\rm{proj}}(t) + f_{\rm{proj}}(t) \frac{F_{\rm{active}, \lambda}}{F_{\rm{clear}, \lambda}} } \\
    & = 1 - \frac{\left( \frac{R_{p}}{R_{\star}}\right)^2_{\lambda}}{1 - f_{\rm{proj}}(t) + f_{\rm{proj}}(t) \frac{F_{\rm{active}, \lambda}}{F_{clear, \lambda}}} \\
    \left( \Hat{\frac{R_{p}}{R_{\star}}} \right)^{2}_{\lambda, t} & = \frac{\left( \frac{R_{p}}{R_{\star}}\right)^2_{\lambda}}{1 - f_{\rm{proj}}(t) (1 - \frac{F_{\rm{active}, \lambda}}{F_{\rm{clear}, \lambda}} )} \quad\cdot
\end{split}
\end{equation*}

\noindent
The correction/ stellar contamination factor between the measured transit depth and the true transit depth of the planet is: 

\begin{equation}\label{contFac}
    \epsilon_{\lambda, t} =\frac{1}{1 -  f_{\rm{proj}}(t) (1 - \frac{F_{\rm{active}, \lambda}}{F_{\rm{clear}, \lambda}} )} \quad\cdot
\end{equation}

\noindent
Rearranging equation \ref{contFac}, we get

\begin{equation}
\begin{split}
    \epsilon_{\lambda, t} & = \frac{1}{1 - f_{\rm{proj}}(t) + f_{\rm{proj}}(t)\frac{F_{\rm{active}, \lambda}}{F_{\rm{clear}, \lambda}}} \\
    & = \frac{F_{\rm{clear}, \lambda}}{ [1 - f_{\rm{proj}}(t)] F_{\rm{clear}, \lambda} + f_{\rm{proj}}(t) F_{\rm{active}, \lambda} } \\
    & = \frac{F_{\rm{clear}, \lambda}}{F_{\rm{out, \lambda}}} \\
    & = \frac{\textrm{Clear stellar spectrum}}{ \textrm{Stellar spectrum at epoch of interest}}
\end{split}
\end{equation}

The above equation can also be useful for constraining the contamination effect without any prior knowledge about the properties of the active regions like its covering fraction or temperature during the planetary transit. However, that requires obtaining a stellar spectrum when no surface inhomogeneities are present. This is quite difficult as some level of inhomogeneities will always be present even during the minima of the magnetic activity of the star.

\section{Corner plot from the stellar surface retrievals}\label{Section 7.2}

\setlength{\tabcolsep}{20pt}
\renewcommand{\arraystretch}{1.5}

\begin{figure*}[]
\centering
    \includegraphics[width=18cm]{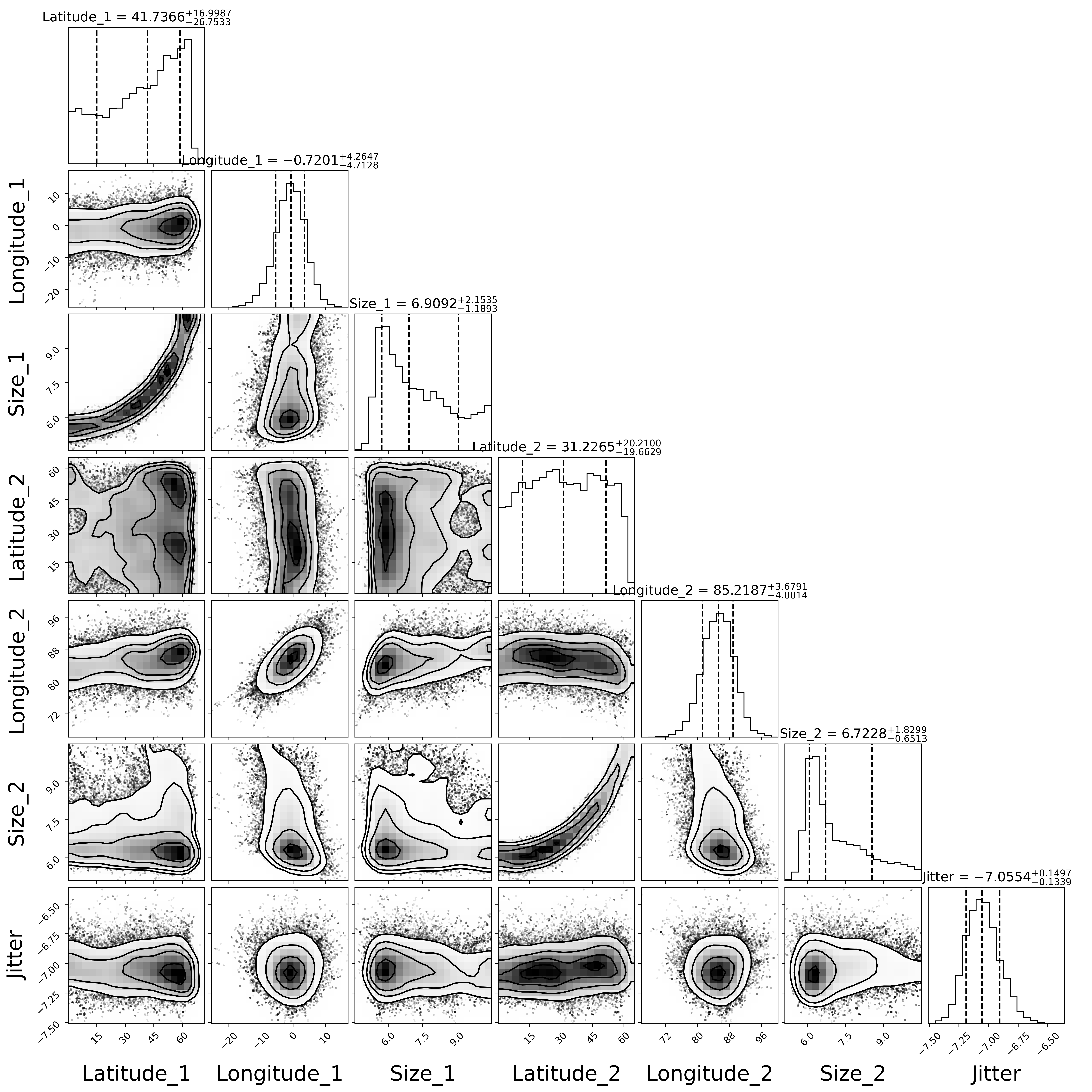}
    \caption{ Posterior distribution of the spot parameters from the stellar surface retrievals of WASP-69 using a 2-spot model. 
    }
    \label{corner-surface-retrivals}
\end{figure*}

\setlength{\tabcolsep}{20pt}
\renewcommand{\arraystretch}{1.5}

\begin{figure*}[]
\centering
    \includegraphics[width=18cm]{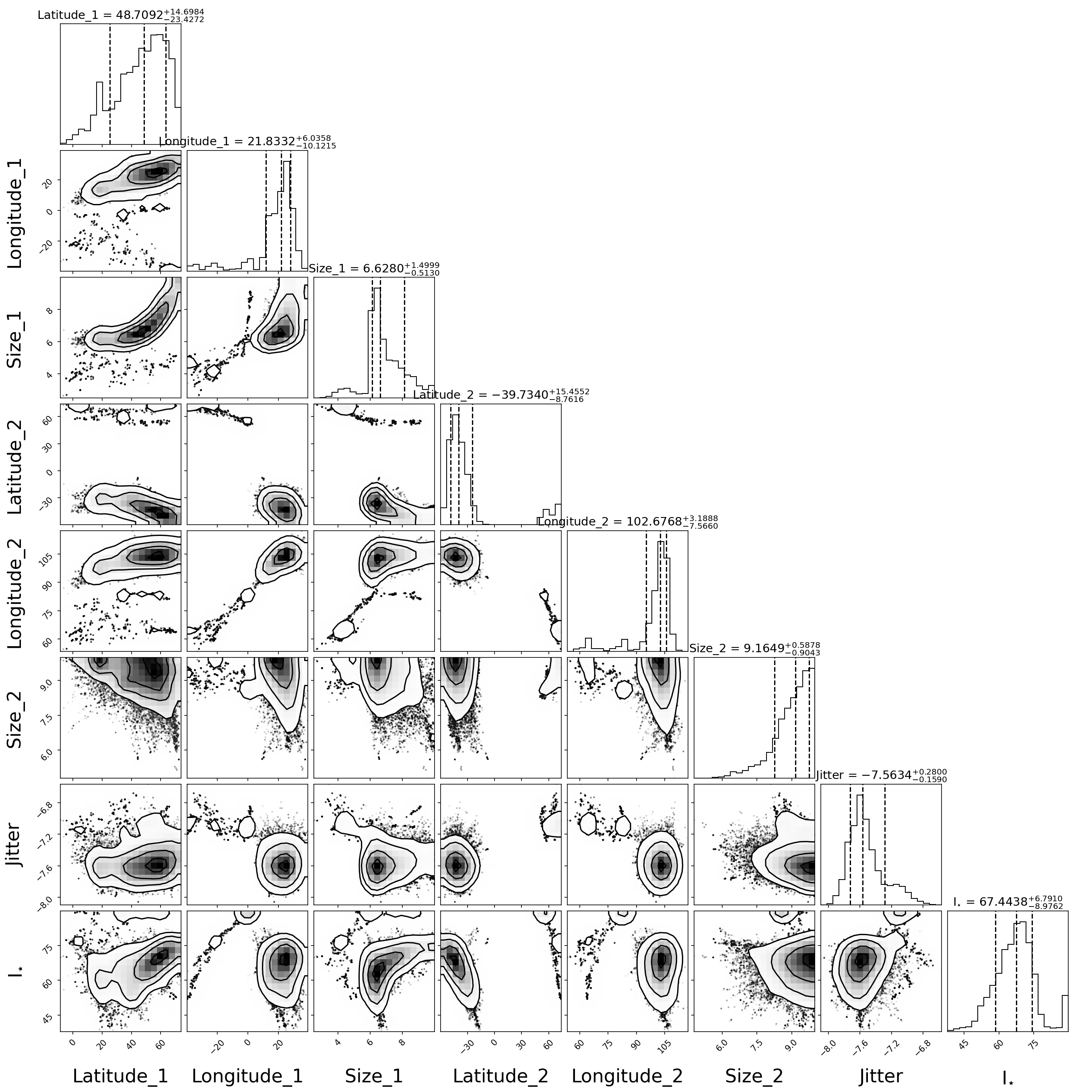}
    \caption{As Fig. \ref{corner-surface-retrivals}, but with stellar inclination ($i_{\star}$) as a free parameter.}
    \label{corner-surface-retrivals_with_inclination}
\end{figure*}

\end{appendix}    

\end{document}